\documentclass[letterpaper,prb,preprint]{revtex4}
\usepackage{epsfig}
\usepackage{subfigure}
\graphicspath{ {./images/} }
\usepackage{bm}
\begin{document}
\title{Li-doped Beryllonitrene for Enhanced Carbon Dioxide Capture}
\author{Andrew Pu}
\affiliation{National Graphene Research and Development Center, Springfield, Virginia 22151, USA}
\author{Xuan Luo}
\affiliation{National Graphene Research and Development Center, Springfield, Virginia 22151, USA}
\date{\today}

\begin{abstract}
\setlength{\parindent}{1cm}
In recent years, the scientific community has given more and more attention to the issue of climate change and global warming, which is largely attributed to the massive quantity of carbon dioxide emissions. Thus, the demand for a carbon dioxide capture material is massive and continuously increasing. In this study, we perform first-principle calculations based on density functional theory to investigate the carbon dioxide capture ability of pristine and doped beryllonitrene. Our results show that carbon dioxide had an adsorption energy of -0.046 eV on pristine beryllonitrene, so it appears that beryllonitrene has extremely weak carbon dioxide adsorption ability. Pristine beryllonitrene could be effectively doped with Lithium atoms, and the resulting Li-doped beryllonitrene had much stronger interactions with carbon dioxide than pristine beryllonitrene. The adsorption energy for carbon dioxide on Li-doped beryllonitrene was -0.408 eV. The adsorption of carbon dioxide on Li-doped beryllonitrene greatly changed the charge density, projected density of states, and band structure of the material, demonstrating that it was strongly adsorbed. This suggests that Li-doping is a viable way to enhance the carbon dioxide capture ability of beryllonitrene and makes it a possible candidate for an effective CO$_2$ capture material.
\end{abstract}

\newpage
\maketitle
\section{Introduction}
Climate change has been an important issue facing the world since the 20th century, and research has shown that this is largely a result of the increase in greenhouse gas emissions\cite{crowley2000}.
The majority of these greenhouse gas emissions are from carbon dioxide (CO$_2$). In 2016, CO$_2$ accounted for about 81.6\% of all manmade greenhouse gas emissions in the US\cite{EPA2017}.
Global warming caused by the increase in greenhouse gas emissions has caused the arctic ice caps to melt, resulting in higher sea levels\cite{bekkers2018}.
Rising temperatures also increase the prevalence of droughts and heatwaves, which are both significant threats to the safety of mankind\cite{benzie2011}.
Thus, the demand for a material to efficiently capture CO$_2$ from the air has been massive.
Currently, there are 3 main methods for CO$_2$ capture: pre-combustion, post-combustion, and oxyfuel\cite{zhou2014}.
When applied to the burning of fossil fuels, pre-combustion and oxyfuel CO$_2$ capture are both performed before CO$_2$ is actually released into the air, while post-combustion CO$_2$ capture refers to the process of capturing CO$_2$ from the resulting gas mixtures.
Thus, post-combustion CO$_2$ capture methods are more applicable to the issue of capturing existing greenhouse gasses from the atmosphere\cite{kanniche2010}.

Previously, several materials and processes for post-combustion CO$_2$ capture have been explored.
Two of the most promising methods of post combustion CO$_2$ capture are chemical absorption and adsorption\cite{yu2012}.
Amine scrubbing, which is one process used for chemical absorption CO$_2$ capture, has been practiced since its patent in 1930\cite{rochelle2009}.
However, chemical absorption processes usually require large equipment, large volumes of solvent, high regeneration energy, and toxic byproducts\cite{haszeldine2009}.
On the other hand, several adsorbents have been researched for their CO$_2$ capture ability\cite{yu2012}.
Many chemical CO$_2$ adsorbents have been studied, including several amine-based adsorbents, but their low CO$_2$ capacity and high cost have reduced their viability\cite{yu2012}.
Meanwhile, several different physical adsorbents have also been explored and research, including porous coal, carbonaceous material, zeolite, fullerenes, and MOFs\cite{yu2012,tawfik2015}.
Many studies have been performed to investigate the possibility of pressure and temperature regulated CO2 capture using physical adsorbents\cite{ishibashi1996,takamura2001}.
However, for all adsorptive CO2 capture methods, the gaseous feed must be treated, and the gaseous mixture usually has to be cooled and dried\cite{metz2005}.
Since the discovery of graphene in 2004, two-dimensional materials have also gathered lots of interest as physical CO$_2$ adsorbents\cite{novoselov2004}.

Two-dimensional materials have a large surface area, atomic-thin structure, and a large surface-to-volume ratio, making them great potential candidates for CO$_2$ capture\cite{shivananju2019}.
Unfortunately, previous experiments have shown that pristine graphene doesn’t demonstrate particularly strong interactions with CO$_2$\cite{wehling2008}.
This has prompted other researchers to investigate the possibility of two-dimensional materials other than graphene for carbon dioxide capture.
Hexagonal boron nitride has a very similar structure to graphene, but it displays greater thermal conductivity, and the nitrogens are believed to be potential active sites for carbon dioxide chemisorption\cite{jiao2011}.
Jiao et al performed density functional theory calculations to investigate this but found that pristine hexagonal boron nitride monolayers were also ineffective for carbon dioxide adsorption.
Silicene and germanene were thought to be potential materials because unlike graphene and boron nitride, they have a buckled hexagonal structure, which gives them greater chemical reactivity\cite{aghaei2018}.
Both pristine silicene and pristine germanene only demonstrated physisorption with CO$_2$\cite{hussain2016,xia2014}.
Monolayer molybdenum disulfide (MoS$_2$) was also investigated due to its unique triple layered buckled structure\cite{yu2015}.
Like the other two-dimensional materials, pristine MoS$_2$ also did not display much adsorption ability for CO$_2$, so it cannot be directly used as an efficient sorbent for CO$_2$\cite{sun2017}.
Several other two-dimensional materials have been investigated for their CO$_2$ capture ability, including aluminum nitride\cite{jia2020}, zinc oxide\cite{rao2015}, gallium nitride\cite{yong2017}, silicon phosphide\cite{fu2021}, silicon carbide\cite{zhang2021}, graphyne\cite{darvishnejad2020,darvishnejad2021,zhou2021}, borophene\cite{luo2020}, and several different carbon nitrides\cite{tan2015,qin2018,he2020electric,he2020novel}.
However, most of these 2D materials are unable to strongly chemisorb CO$_2$ while in their pristine form.
On the other hand, other studies have found that dopants and defects were able to greatly increase the adsorption capabilities of many of these 2D materials.
Li et al and Chandra et al both found that graphene's CO$_2$ capture ability was greatly increased after nitrogen doping\cite{li2017,chandra2012}.
Boron nitride was shown to have increased CO$_2$ adsorption behavior after fluorine doping and boron vacancies\cite{jiao2011,liu2021}.
Silicene and Germanene both displayed much greater CO$_2$ capture ability after Li-functionalization\cite{aghaei2018,zhu2016}, and MoS$_2$ monolayers with S vacancies also demonstrated much better adsorption ability\cite{yu2015}.
Recently, Bykov et al discovered a new two-dimensional material known as beryllonitrene, which could be a potential CO$_2$ capture material due to its unusual structure and variety of binding sites\cite{bykov2021}.

Beryllonitrene is a monolayer form of BeN$_4$ that was created by performing high-pressure synthesis followed by decompression on triclinic BeN$_4$\cite{bykov2021}.
Not much research has been conducted regarding its practical applications, including in the field of gas capture and gas sensing.
Beryllonitrene is similar to the previously stated two-dimensional materials in that it's structure is close to a honeycomb structure, and it has a hexagonal unit cell\cite{bafekry2021}.
Each unit cell consists of one beryllium atom and four nitrogen atoms, so the material is mostly composed of nitrogen\cite{bafekry2021}.
Like in boron nitride, these nitrogens could act as potential carbon dioxide adsorption sites\cite{jiao2011}.
Beryllonitrene has been shown to be both dynamically and thermodynamically stable\cite{bykov2021}.
The monolayer exhibits semimetallic behavior\cite{bafekry2021} and has a large surface area, similar to other two-dimensional materials, making it a good candidate for CO$_2$ adsorption.
Due to the unique properties and lack of research for beryllonitrene, it is important  to investigate the potential of beryllonitrene for CO$_2$ capture.

In this study, we use density functional theory to calculate the adsorption of CO$_2$ molecules onto the surface of pristine beryllonitrene and Li-doped beryllonitrene in order to determine whether they have the potential to be effectively used as a CO$_2$ capture device.

\section{Methods}

\subsection{Computational Details}
First-principle calculations using density functional theory (DFT)\cite{kohn1965,hohenberg1964} as implemented in the ABINIT program\cite{gonze2020,romero2020,gonze2016,gonze2009,gonze2005,gonze2002} were performed using the generalized gradient approximation (GGA) with the  Perdew-Burke-Ernzerhof (PBE) formalism as the exchange-correlation functional\cite{perdew1996}. Projector augmented-wave (PAW)\cite{blochl1994,torrent2008} pseudopotentials were selected, which were generated by ATOMPAW code\cite{holzwarth2001}.
The electron configurations and radius cutoffs used for generating the PAW pseudopotentials are listed in Table \ref{electron}.

\begin{table}[hpt]
\caption{Electron configurations and radial cut offs for all elements used to generate PAW pseudopotentials}
\begin{tabular}{|c|c|c|}
\hline
Element&Electron Configuration&r$_{cut}$ (bohr)\\
\hline
Beryllium (Be)&1s2 2s2&1.6\\
\hline
Nitrogen (N)&[He] 2s2 2p3&1.2\\
\hline
Aluminum (Al)&[Ne] 3s2 3p1&1.9\\
\hline
Phosphorus (P)&[Ne] 3s2 3p3&1.9\\
\hline
Sulfur (S)&[Ne] 3s2 3p4&1.9\\
\hline
\end{tabular}
\label{electron}
\end{table}

Convergence calculations were performed to determine appropriate converged values for the kinetic energy cutoff, k point mesh, and vacuum. The values were considered converged when the difference in total energy between two data sets was less than $1.0\times10^{-4}$ Ha two times consecutively. During the convergence calculations, the self-consistent field (SCF) total energy calculations were considered complete when the total energy difference was less than $1.0\times10^{-10}$ Ha twice. For the relaxation of the lattice parameters and atomic structure, the Broyden-Fletcher-Goldfarb-Shanno (BFGS) method\cite{Head1985} was used. During the BFGS relaxation calculations, the SCF iteration was terminated when the total difference in forces was less than $2.0\times10^{-5}$ Ha/bohr two times consecutively. The BFGS relaxation calculations were considered complete when all of the forces were less than $2.0\times10^{-4}$ Ha/bohr.

\subsection{CO$_2$ Adsorbed on Pure Beryllonitrene}

Convergence calculations were performed on a $1\times1$ unit cell of beryllonitrene, shown in Figure \ref{brillouin}(a), and then converged values were used for relaxation of the atomic structure. Convergence and relaxation calculations were also performed on a single carbon dioxide molecule. Calculations for the adsorption of carbon dioxide on beryllonitrene were performed using a $2\times2$ surpercell of beryllonitrene.

The adsorption energy of CO$_2$, $E_{ads}$, on each material was calculated by
\begin{equation} \label{eq1}
E_{ads}=E_{ber+CO_2}-E_{ber}-E_{CO_2}
\end{equation}
\noindent
where $E_{CO_2}$, $E_{ber}$, and $E_{ber+CO_2}$ represent the total energy for isolated carbon dioxide, pristine or doped beryllonitrene, and the carbon dioxide and beryllonitrene complex, respectively.

\subsection{CO$_2$ Adsorbed on Doped Beryllonitrene}

Convergence and relaxation calculations were also performed for $2\times2$ supercells of doped beryllonitrene. Calculations for the adsorption of carbon dioxide on the doped beryllonitrene were also performed, and the adsorption energy was found using Equation \ref{eq1}.

The defect formation energy, $E_{def}$, for each doped beryllonitrene monolayer is defined by
\begin{equation} \label{eq2}
E_{def}=E_{ber+dop}-E_{ber}-E_{dop}
\end{equation}
where $E_{ber}$, $E_{dop}$, and $E_{ber+dop}$ represent the total energy for pristine beryllonitrene, the interstitial dopant (Li, Ca, or Al) in its crystalline structure, and the doped beryllonitrene, respectively.

\subsection{Electronic Structure}

The charge transfer, $\Delta\rho(\tau)$, involved in the adsorption of CO$_2$ was calculated by
\begin{equation} \label{eq3}
\Delta\rho(\tau)=\rho_{ber+CO_2}(\tau)-\rho_{ber}(\tau)-\rho_{CO_2}(\tau)
\end{equation}
where $\rho_{CO_2}(\tau)$, $\rho_{ber}(\tau)$, and $\rho_{ber+CO_2}(\tau)$ represent the charge density of carbon dioxide, beryllonitrene, and the carbon dioxide and beryllonitrene complex, respectively.

\begin{figure}[h!]
\includegraphics [width=10cm]{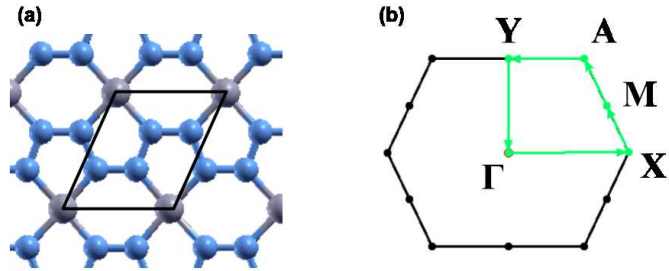}
\caption{(a) Atomic structure of pristine beryllonitrene. The black frame indicates a $1\times1$ cell (b) First Brillouin zone for beryllonitrene with high symmetry points $\Gamma$, X, M, A, and Y\cite{mortazavi2021}.}
\label{brillouin}
\end{figure}

The band structure was calculated for each beryllonitrene monolayer and each complex using $\Gamma$, X, M, A, and Y as the high symmetry k points, as shown in Figure \ref{brillouin}(b).
The tetrahedron method was used to plot the projected density of states (PDOS) for the different monolayers and complexes. The PDOS graphs use the 2s orbital for lithium and beryllium, as well as the 2p orbitals for carbon, nitrogen, and oxygen.

\section{Results \& Discussion}

In this study, we first optimized the structure of beryllonitrene, Li-doped beryllonitrene, and a single carbon dioxide molecule. We then calculated the adsorption energy of carbon dioxide on each of the monolayers. The carbon dioxide was considered chemisorbed when the absolute value of the adsorption energy was greater than the chemisorption threshold of -0.30 eV, which is equivalent to -0.011 Ha\cite{jia2020,tawfik2015}.

\subsection{Atomic Structure}

The optimized atomic structures for the materials were found using BFGS relaxation calculations, which gives the molecules free range of motion and any bonds formed were chemical bonds\cite{Head1985}.

\subsubsection{Pristine Beryllonitrene and Carbon Dioxide Structure}

\begin{table}[ht]
    \centering
    \caption{Comparison between calculated and previous theoretical values for the lattice parameters (a, b) of beryllonitrene, the bond lengths of the Be-N, N$_1$-N$_2$, and N$_2$-N$_2$ bond in beryllonitrene, and the C-O bond in carbon dioxide (Atom labels shown in Figure \ref{beryllonitrene})}
    \begin{tabular}{|c|c|c|c|c|c|c|}
         \hline
         &a&b&Be-N&N$_1$-N$_2$&N$_2$-N$_2$&C-O\\
         \hline
         Calculated (\AA)&3.660&4.270&1.747&1.341&1.337&1.172\\
         \hline
         Theoretical (\AA)&3.66\cite{mortazavi2021}&4.27\cite{mortazavi2021}&1.748\cite{mortazavi2021}&1.343\cite{mortazavi2021}&1.338\cite{mortazavi2021}&1.161\cite{klotz2004}\\
         \hline
    \end{tabular}
    \label{past}
\end{table}

\begin{figure}[h!]
\includegraphics [width=10cm]{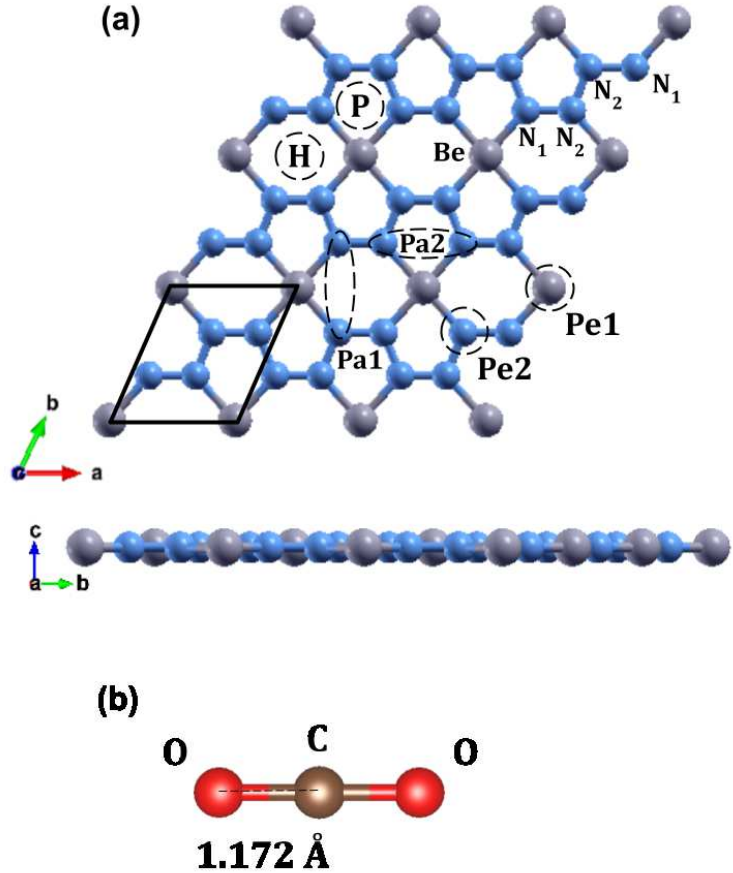}
\caption{(a) Atomic structure of a $3\times3$ supercell of monolayer beryllonitrene (grey atoms are beryllium, and blue atoms are nitrogen). On the beryllonitrene supercell, the CO$_2$ adsorption sites are labeled Pa1, Pa2, Pe1, and Pe2, and the interstitial doping sites are labeled as H and P. Sites Pa1 and Pa2 are with the CO$_2$ parallel to the beryllonitrene monolayer, while sites Pe1 and Pe2 are with the CO$_2$ perpendicular to the beryllonitrene monolayer. Site H is located above the Be$_2$N$_4$ hexagon while site P is located above the BeN$_4$ pentagon. (b) Optimized atomic structure of a single carbon dioxide molecule (red atoms are oxygen, and brown atoms are carbon).}
\label{beryllonitrene}
\end{figure}

Beryllonitrene consists of chains of nitrogen atoms held together by beryllium atoms.
The atomic structure is comprised of BeN$_4$ pentagons and Be$_2$N$_4$ hexagons, as shown in Figure \ref{beryllonitrene}(a).
Table \ref{past} presents the structural properties of the relaxed beryllonitrene, as well as the bond length of carbon dioxide, and compares them with previous theoretical results.
The unit cell of pristine beryllonitrene was optimized and found to have lattice parameters of $3.66$ and $4.27$ \AA.
The hexagonal primitive cell has a $64.64^{\circ}$ angle between the lattice vectors, which is consistent with previous results\cite{mortazavi2021}.
The beryllium-nitrogen bond lengths were found to be approximately $1.75$ \AA, and the nitrogen-nitrogen bond lengths were all around $1.34$ \AA, which is also consistent with previous results\cite{bafekry2021}.
The Be-N$_1$-N$_2$ bond angle is approximately $131.56^{\circ}$, and the N$_1$-N$_2$-N$_2$ bond angle is approximately $111.49^{\circ}$.
A carbon dioxide molecule was also structurally optimized and found to have a linear structure with bond lengths of $1.17$ \AA,\ as shown in Figure \ref{beryllonitrene}(b).

\subsubsection{Carbon Dioxide Adsorption on Pristine Beryllonitrene}

In order to find the most optimal binding site for carbon dioxide, we tested the four beryllonitrene adsorption sites shown in Figure \ref{beryllonitrene}(a): Pa1, Pa2, Pe1, and Pe2.

Full relaxation calculations were performed for each configuration of beryllonitrene and carbon dioxide.
The relaxed values were used to perform total energy calculations for the relaxed configuration.
For the first two configurations, Pa1 and Pa2, we placed the carbon dioxide molecule horizontally above the beryllonitrene such that the oxygen atoms of carbon dioxide were above the nitrogen atoms in beryllonitrene, as shown in Figure \ref{beryllonitrene}(a). Position Pa1 for CO$_2$ is above the BeN$_4$ pentagon and parallel to the a axis. Position Pa2 for CO$_2$ is above the Be$_2$N$_4$ hexagon and perpendicular to the a axis.

In the next two configurations, Pe1 and Pe2, the carbon dioxide was placed perpendicular to the plane of the two-dimensional beryllonitrene, as shown in Figure \ref{beryllonitrene}(a). Position Pe1 for CO$_2$ is above a beryllium atom, and Position Pe2 for CO$_2$ is above a nitrogen atom.

\begin{figure}[h]
\includegraphics[width=10cm]{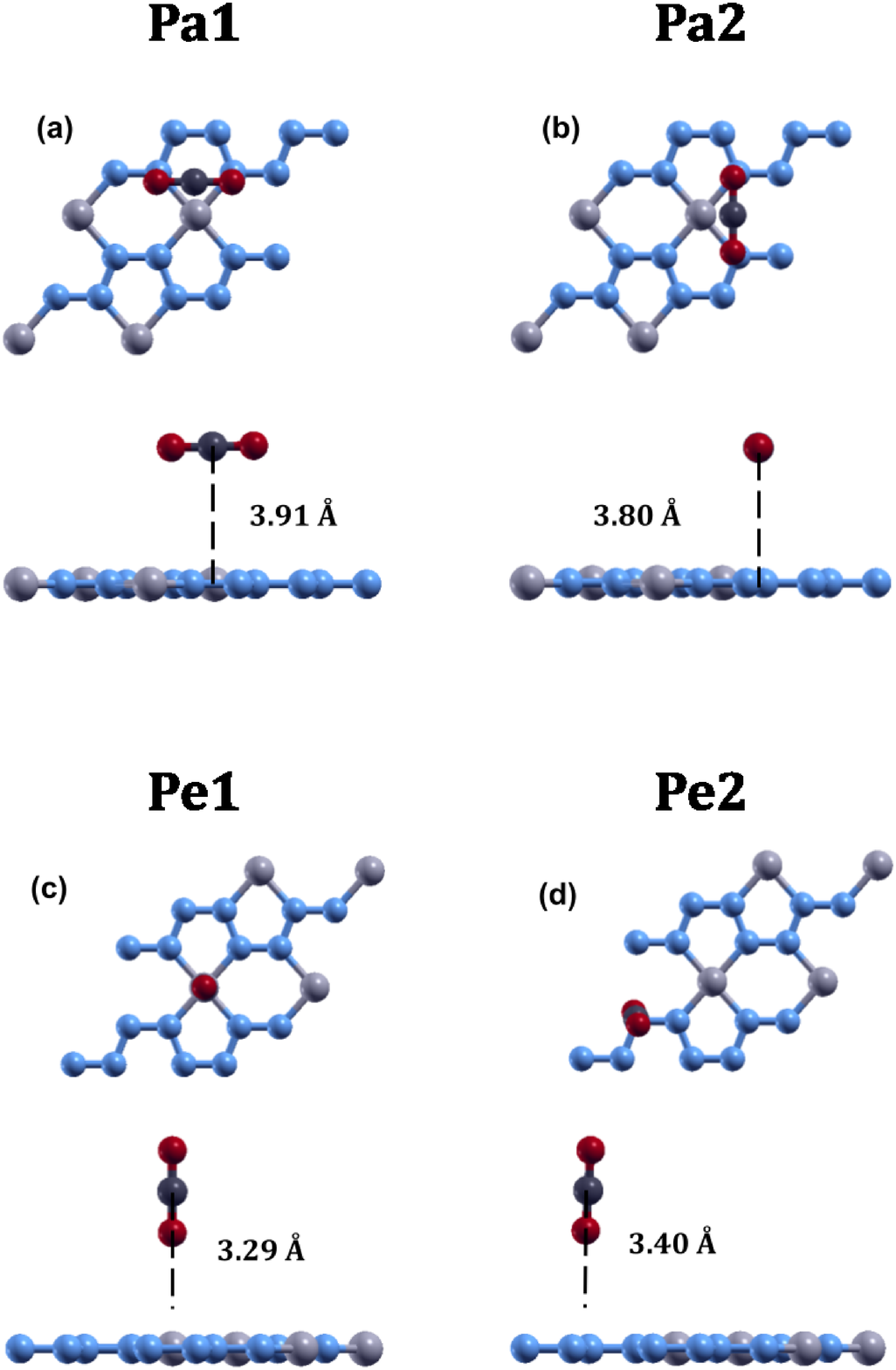}
\caption{Optimized atomic configuration for the adsorption of carbon dioxide on pristine beryllonitrene at (a) position Pa1 (Parallel to BeN$_4$ \& above Be$_2$N$_4$ hexagon), (b) position Pa2 (Parallel to BeN$_4$ \& above BeN$_4$ pentagon), (c) position Pe1 (Perpendicular to BeN$_4$ \& above beryllium), and (d) position Pe2 (Perpendicular to BeN$_4$ \& above nitrogen) after relaxation. The blue atoms are nitrogen, the grey atoms are beryllium, the red atoms are oxygen, and the dark grey atoms are carbon.}
\label{adsorption}
\end{figure}

\begin{table}[h]
    \centering
    \caption{Adsorption energy (E$_{ads}$) for CO$_2$, distance between CO$_2$ and beryllonitrene (d), and CO$_2$ bond angle (\angle{O-C-O}) after adsorption onto pristine beryllonitrene at positions Pa1, Pa2, Pe1, and Pe2 as shown in Figure \ref{adsorption}}
    \begin{tabular}{|c|c|c|c|}
        \hline
         Position&E$_{ads}$ (Ha)&d (\AA)&\angle{O-C-O} ($^{\circ})$\\
         \hline
         Pa1&$-0.0016$&$3.91$&180\\
         \hline
         Pa2&$-0.0017$&$3.80$&180\\
         \hline
         Pe1&$0.0000$&$3.29$&180\\
         \hline
         Pe2&$0.0000$&$3.40$&180\\
         \hline
    \end{tabular}
    \label{CO2}
\end{table}

The relaxed atomic structure for all four configurations is shown in Figure \ref{adsorption}. In all 4 configurations, the bond angle of carbon dioxide stayed at around $180^\circ$. The most optimal position for carbon dioxide adsorption on beryllonitrene was found to be position Pa2, which was 3.80 \AA\ above the BeN$_4$ pentagon and perpendicular to the a axis , as shown in Figure \ref{adsorption}(b). This configuration has an adsorption energy of -0.0017 Hartree, which is smaller than the chemisorption threshold of -0.011 Ha. For comparison, the adsorption energy for carbon dioxide on pristine boron nitride is -0.0018 Ha\cite{jiao2011}. This suggests that the carbon dioxide molecule is only physically adsorbed onto the surface of pristine beryllonitrene. Since all of the configurations of carbon dioxide on pristine beryllonitrene had adsorption energies smaller than -0.011 Hartree\cite{jia2020,tawfik2015}, this means that carbon dioxide cannot spontaneously chemisorb onto the beryllonitrene, and pristine beryllonitrene is not a viable adsorbent for CO$_2$ capture.

\subsubsection{Adsorption of Dopants onto Beryllonitrene Monolayer}

We tested the adsorption of Li, Ca, and Al onto beryllonitrene as dopants. In previous experiments, it was found that there are 2 optimal binding sites for ions on beryllonitrene: Site H (above the Be$_2$N$_4$ hexagon) and Site P (above the BeN$_4$ pentagon)\cite{mortazavi2021} as shown in Figure \ref{beryllonitrene}(a). Relaxation calculations were performed for beryllonitrene with the dopants at both H and P.

\begin{figure}[h]
\includegraphics[width=10cm]{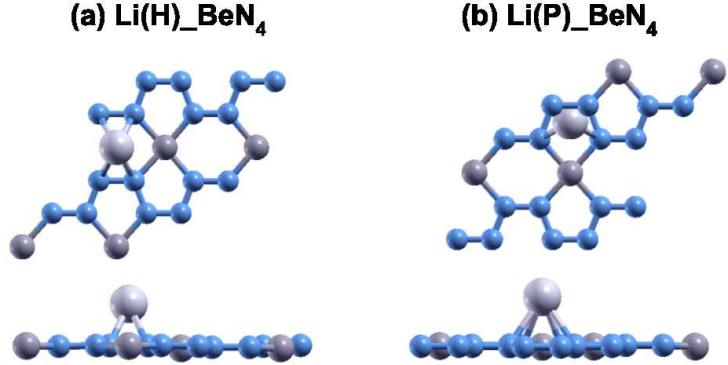}
\caption{Optimized atomic structure of beryllonitrene $2\times2$ supercell doped with Li at (a) H (above the center of the Be$_2$N$_4$ hexagon) and at (b) P (above the center of the BeN$_4$ pentagon) after relaxation. The blue atoms are nitrogen, the dark grey atoms are beryllium, and the light grey atoms are lithium.}
\label{lidope}
\end{figure}

\begin{figure}[h]
\caption{Optimized atomic structure of beryllonitrene doped with Ca at (a) H and (b) P, as well as beryllonitrene doped with Al at (c) H and (d) P after relaxation. Positions H (above the center of the Be$_2$N$_4$ hexagon) and P (above the center of the BeN$_4$ pentagon) are shown in Figure \ref{beryllonitrene}(a). The blue atoms are nitrogen, the dark grey atoms are beryllium, the turquoise atoms are calcium, and the light grey atoms are aluminum.}
\includegraphics[width=10cm]{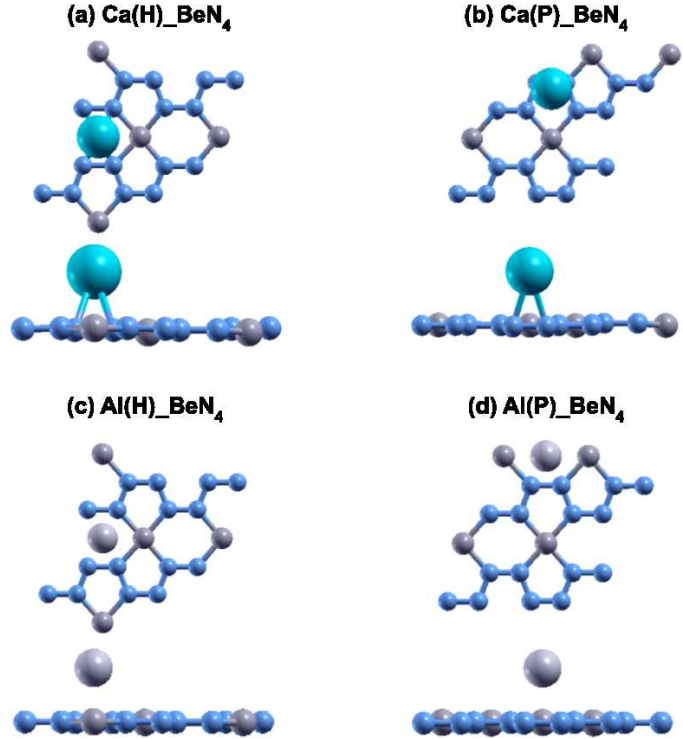}
\label{caaldope}
\end{figure}

The relaxed atomic structures for each of the doped beryllonitrene supercells are shown in Figure \ref{lidope} \& \ref{caaldope}.
When placed at H, all three of the dopants appeared to stay in the same spot. When placed on site P, the dopants tended to move perpendicular to the a axis towards site H. When Al was placed on site P, the Al atom actually moved all the way to site H.
The total energy for each of the dopant-beryllonitrene complexes was greater when the dopant was placed at H rather than P.
Adsorption of Ca and Al on both sites resulted in positive adsorption energies, which suggests that Ca and Al atoms would prefer to form metallic clusters rather than adsorb to the surface of beryllonitrene\cite{mortazavi2021} at these two sites. On the other hand, Li has negative adsorption energies at both sites, so it is likely to adsorb to the surface of beryllonitrene. On site H, Li had an adsorption energy of -0.0192 Ha, while Li on site P had an adsorption of energy of -0.0121 Ha.
Both of these adsorption energies are greater than -0.011 Hartree, which is considered the chemisorption threshold\cite{jia2020,tawfik2015}, so Li is considered to be chemically adsorbed onto beryllonitrene on both of these sites. Since Li had greater adsorption energy on site 1 than on site 2, Li(H)\_BeN$_4$ is more stable than Li(P)\_BeN$_4$ (shown in Figure \ref{lidope}).

\begin{table}[h]
    \centering
    \caption{Defect formation energies (E$_{def}$) and height from beryllonitrene (h) for the following dopants on beryllonitrene: Lithium at site H and site P (Figure \ref{lidope}), as well as Calcium and Aluminum at site H and site P (Figure \ref{caaldope})}
    \begin{tabular}{|c|c|c|}
    \hline
         Complex&E$_{def}$ (Ha)&h (\AA)\\
    \hline
         Li\_H\_BeN$_4$&-0.0192&1.49\\
    \hline
         Li\_P\_BeN$_4$&-0.0121&1.76\\
    \hline
         Ca\_H\_BeN$_4$&0.0498&2.05\\
    \hline
         Ca\_P\_BeN$_4$&0.0489&2.26\\
    \hline
         Al\_H\_BeN$_4$&0.0141&2.00\\
    \hline
         Al\_P\_BeN$_4$&0.0142&2.01\\
    \hline
    \end{tabular}
    \label{edoping}
\end{table}

\subsubsection{Carbon Dioxide Adsorption on Doped Beryllonitrene}

\begin{figure}[h]
\includegraphics[width=10cm]{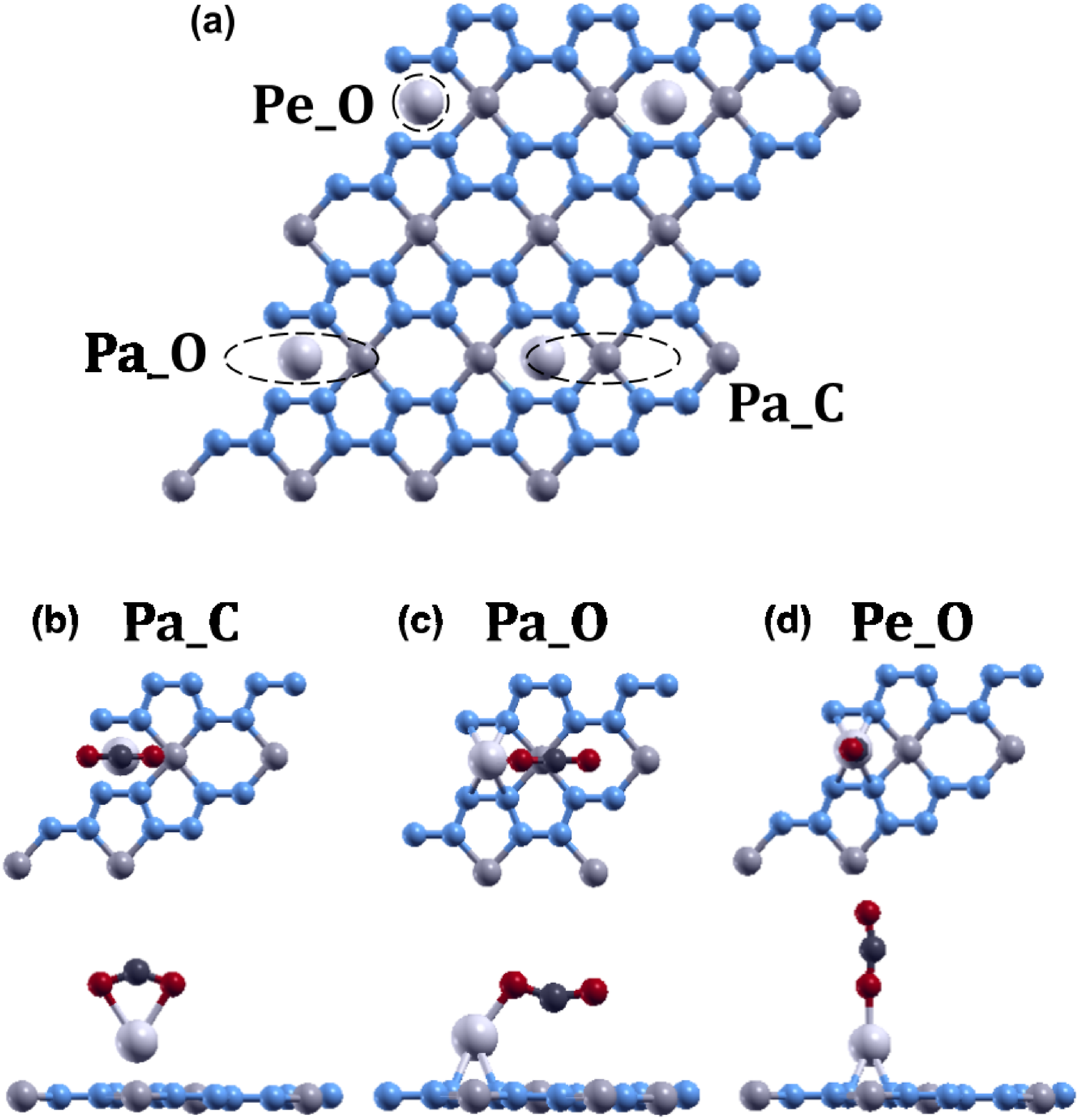}
\caption{(a) Three positions for carbon dioxide adsorption on Li-doped beryllonitrene. Optimized structure of carbon dioxide placed onto Li-doped beryllonitrene at positions (b) Pa\_C (Parallel to BeN$_4$ \& carbon atom above lithium) (c) Pa\_O (Parallel to BeN$_4$ \& oxygen atom above lithium) and (d) Pe\_O (Perpendicular to BeN$_4$ \& above lithium). The blue atoms are nitrogen, the medium grey atoms are beryllium, the light grey atoms are lithium, the red atoms are oxygen, and the dark grey atoms are carbon.}
\label{doped}
\end{figure}

Carbon dioxide adsorption calculations were performed on Li(H)\_BeN$_4$ as it was found to be the most stable doped form of beryllonitrene that was tested, as shown in Figure \ref{lidope}(a). Three sites on Li-doped beryllonitrene were tested for CO$_2$ adsorption as shown in Figure \ref{doped}.
Position Pa\_C is horizontal with the carbon atom directly above the lithium atom. Position Pa\_O is horizontal with an oxygen atom directly above the lithium atom. Position Pe\_O is vertical directly above the lithium atom.

Of these three positions, CO$_2$ molecules only chemisorbed to the Li-doped beryllonitrene when placed in positions Pa\_O and Pe\_O.
The adsorption energy for CO$_2$ at position Pa\_C is 0.0002 Ha, which is positive, suggesting that this is not a viable position for carbon dioxide to adsorb onto beryllonitrene.
In this position, the oxygen atoms were pulled closer to the lithium atom than the carbon atom, resulting in a bond angle of 142.192$^{\circ}$.
The adsorption energy for CO$_2$ at position Pa\_O is -0.0155 Ha, while the adsorption energy for CO$_2$ at position Pe\_O is -0.0145 Ha.
Both of these adsorption energies are greater than the chemisorption threshold of 0.011 Ha, so the CO$_2$ molecule is chemisorbed to Li-doped beryllonitrene at these two positions\cite{jia2020,tawfik2015}.
This suggests that the oxygen atoms in the CO$_2$ molecules are attracted to the Li dopant, so CO$_2$ molecules are more likely to adsorb to Li-doped beryllonitrene with an oxygen atom pointed towards the lithium atom.

\begin{table}[h]
    \centering
    \begin{tabular}{|c|c|c|c|}
        \hline
         Position&E$_{ads}$ (Ha)&d (\AA)&\angle{O-C-O} ($^{\circ}$)\\
        \hline
         Pa\_C&0.0002&2.15&142\\
        \hline
         Pa\_O&-0.0155&1.91&153\\
        \hline
         Pe\_O&-0.0145&1.90&165\\
        \hline
    \end{tabular}
    \caption{Adsorption energies (E$_{ads}$) for CO$_2$, distance between CO$_2$ and lithium (d), and CO$_2$ bond angle (\angle{O-C-O}) on Li-doped beryllonitrene at positions Pa\_C, Pa\_O, and Pe\_O}
    \label{eadsorption}
\end{table}

Since CO$_2$ had the greatest adsorption energy when placed in position Pa\_O, this configuration appears to be the most stable configuration, so it was used for further calculations.

\subsection{Charge Transfer}
\begin{figure}[h]
\includegraphics[width=10cm]{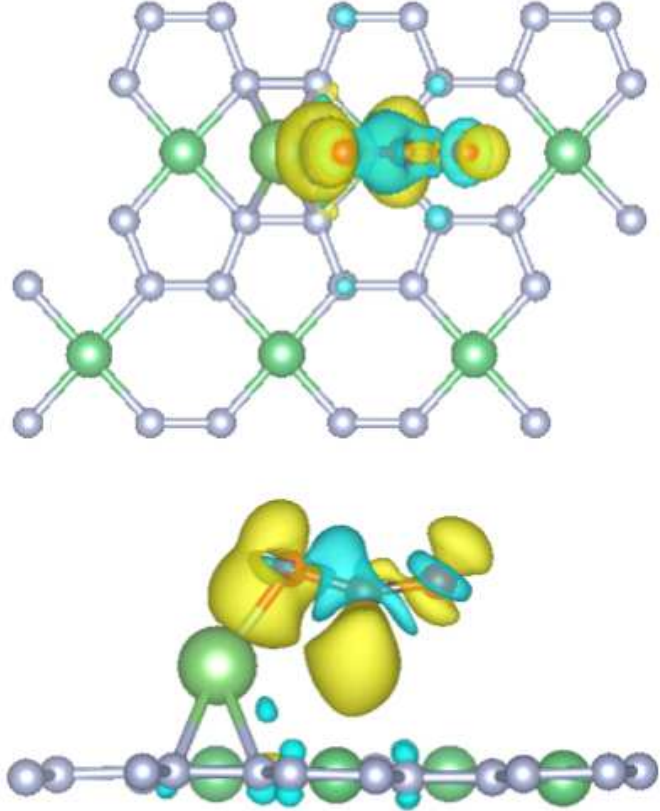}
\caption{Top and side view of charge transfer isosurface for adsorption of CO$_2$ on Li-doped beryllonitrene. Yellow areas are charge gained and blue areas are charge lost.}
\label{charge}
\end{figure}

The charge transfer diagram was plotted for the adsorption of CO$_2$ onto Li-doped beryllonitrene at site Pa\_O, as shown in Figure \ref{charge}. There appears to be a transfer of electrons away from the carbon atom and towards the oxygen atoms, particularly the atom facing the lithium dopant. This makes sense because oxygen is much more electronegative than carbon and lithium, so it has a much stronger attraction towards the electrons. The existence of the charge transfer indicates that a chemical bond was formed between the carbon dioxide molecule and the Li-doped beryllonitrene.

\subsection{PDOS}

\begin{figure}[h]
\includegraphics[width=8cm]{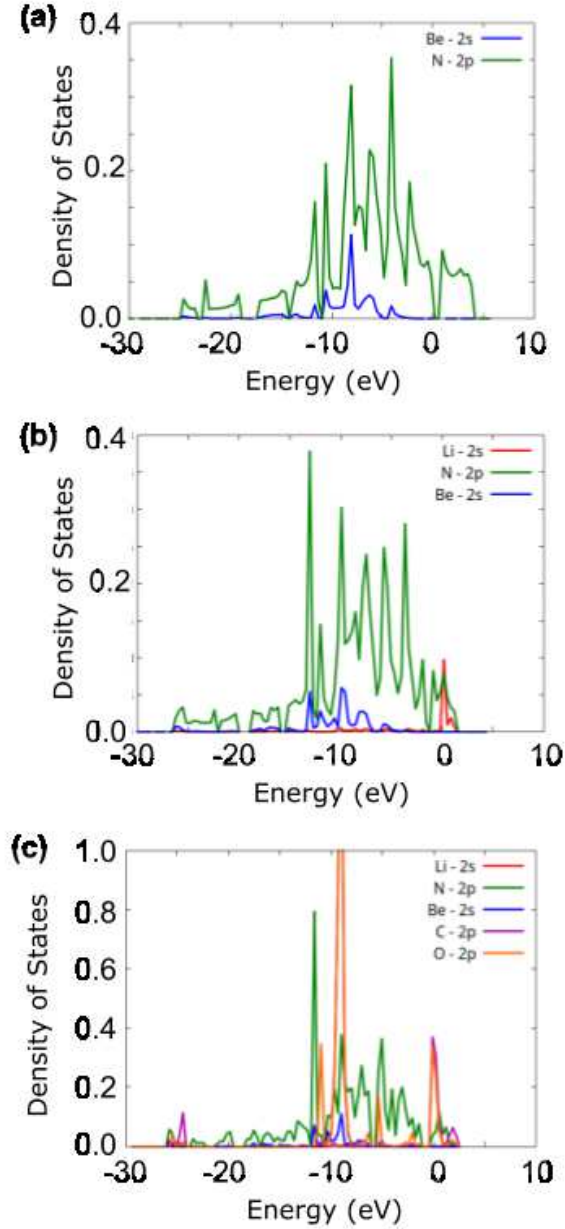}
\caption{Projected density of states (PDOS) for (a) pristine beryllonitrene, (b) Li-doped beryllonitrene, and (c) Li-doped beryllonitrene with carbon dioxide adsorbed.}
\label{pdos}
\end{figure}

To further understand the adsorption of CO$_2$ onto Li-doped beryllonitrene, the projected density of states (PDOS) was plotted for pristine beryllonitrene, Li-doped beryllonitrene with Li at position H as shown in Figure \ref{beryllonitrene}(a), and Li-doped beryllonitrene with carbon dioxide at Pa\_O as shown in Figure \ref{doped}(a).

For pristine beryllonitrene, the 2p orbital of nitrogen (green line) and the 2s orbital for beryllium (blue line) were plotted, as shown in Figure \ref{pdos}(a). There appears to be significant overlap between the 2p orbital of nitrogen and the 2s orbital of beryllium in the range from about -10 to -5 eV. This overlap indicates that the beryllium and nitrogen atoms in the beryllonitrene monolayer are chemically bonded, which means that the beryllonitrene monolayer is a plausible monolayer.

For Li-doped beryllonitrene, the 2s orbital of lithium (red line) was also plotted alongside nitrogen and beryllium, as shown in Figure \ref{pdos}(b). The addition of the lithium dopant causes the PDOS to shift towards the left (towards negative energy). The PDOS for the 2s orbital of lithium (red line) has a major peak at around 0 eV, which overlaps with the PDOS spectra of nitrogen, indicating that a chemical bond has formed between the lithium dopant and the nitrogen in beryllonitrene. Lithium does not demonstrate much overlap with beryllium. This matches with the relaxed atomic structure because it shows that lithium was bonded to beryllonitrene via the four nitrogen atoms in the Be$_2$N$_4$ hexagon, as shown in Figure \ref{lidope}(a).

The PDOS spectra was also plotted for the Li-doped beryllonitrene with a carbon dioxide molecule adsorbed at Pa\_O, as shown in Figure \ref{pdos}(c). In addition to the spectra for lithium, nitrogen, and beryllium, the 2p orbitals of oxygen (orange line) and carbon (purple line) were also plotted.
Lithium's largest peaks were located at around 2 and -10 eV, and had significant overlap with the carbon and oxygen molecules at these peaks. This suggests that the carbon dioxide molecule was chemically bonded to the lithium atom. The PDOS spectra of nitrogen also overlapped with lithium at these peaks, indicating that the lithium dopant was still bound to the beryllonitrene. These results show that carbon dioxide was able to successfully chemically bond to Li-doped beryllonitrene.

\subsection{Band Structure}

\begin{figure}[h]
\includegraphics[width=8cm]{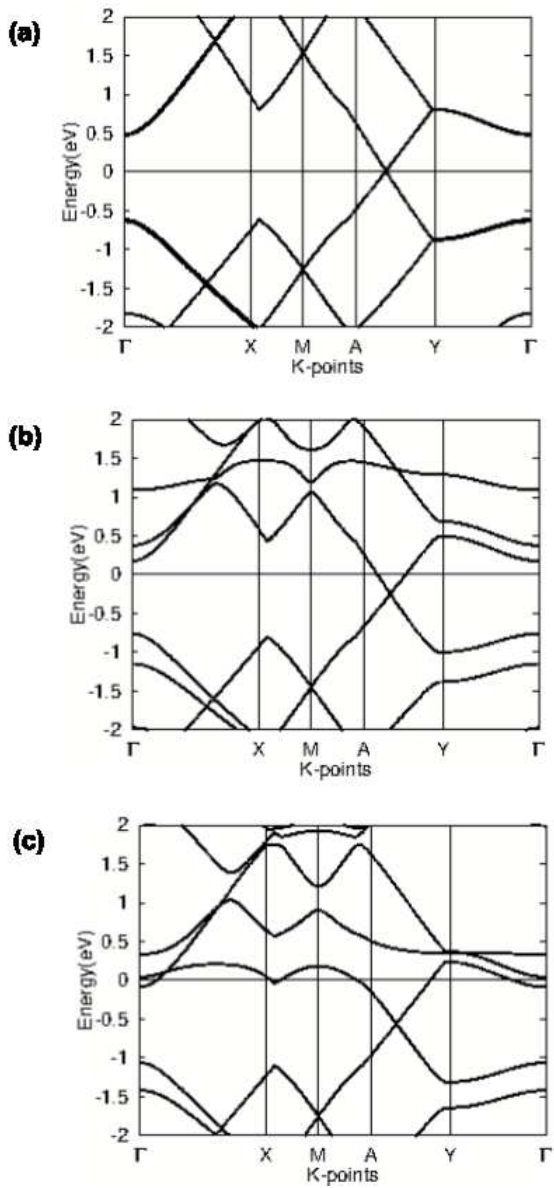}
\caption{Band structure for (a) pristine beryllonitrene, (b) Li-doped beryllonitrene, and (c) Li-doped beryllonitrene with carbon dioxide adsorbed.}
\label{band}
\end{figure}

In order to better understand the electronic properties of the materials, the band structure was plotted for pristine beryllonitrene, Li-doped beryllonitrene, and Li-doped beryllonitrene with CO$_2$ adsorbed at Pa\_O. Figure \ref{band} shows the calculated band structures for these materials.

The calculated band structure for a 2x2 supercell of pristine beryllonitrene is in Figure \ref{band}(a), and it shows that beryllonitrene is a zero-gap semimetal with a Dirac-point between the A and Y k-points, which is consistent with previous results\cite{bafekry2021}.
The band structure for Li-doped beryllonitrene shown in Figure \ref{band}(b) shows that after the lithium dopant is added to the beryllonitrene, the half-filled band from lithium lies entirely above the Fermi level. This means that lithium doping changes beryllonitrene from a semimetal to a metal.
Figure \ref{band}(c) shows the band structure of Li-doped beryllonitrene after carbon dioxide is adsorbed at site Pa\_O. The two bands above the Fermi level appear to have been changed significantly, indicating that the adsorption of CO$_2$ changes the electronic structure of Li-doped beryllonitrene.

Since Li-doped beryllonitrene exhibits metallic behavior, it could potentially be used as a charge-modulated CO$_2$ capture material. Several other metallic materials have already been theoretically tested for charge-modulated CO$_2$ capture\cite{tan2017,zhang2021}.

\section{Conclusion}

In summary, we used first principle calculations to determine the carbon dioxide capture ability of pristine beryllonitrene and Li-doped beryllonitrene. The atomic structure, adsorption energy, charge transfer, projected density of states, and band structure were analyzed to determine the viability of carbon dioxide adsorption. We found that pristine beryllonitrene demonstrated very weak attraction towards carbon dioxide. In order to enhance the carbon dioxide adsorption ability, we doped beryllonitrene with three different dopants: lithium, calcium, and aluminum. Calcium and aluminum were unable to spontaneously adsorb, but lithium was able to strongly chemisorb onto pristine beryllonitrene, producing Li-doped beryllonitrene. Lithium doping appeared to greatly enhanced the carbon dioxide capture ability of beryllonitrene since carbon dioxide had an adsorption energy of -0.0155 Ha on Li-doped beryllonitrene, meaning that the CO$_2$ molecule was chemisorbed. The charge transfer, projected density of states, and band structure diagrams show that the adsorption of carbon dioxide greatly change the electronic structure of the Li-doped beryllonitrene, suggesting that carbon dioxide is strongly chemisorbed onto the surface. These results demonstrate that Li-doped beryllonitrene is a possible candidate for application in CO$_2$ capture.

In our study, we did not consider whether or not the adsorption of carbon dioxide on these beryllonitrene monolayers was reversible. Typically, a certain amount of energy is required in order for the carbon dioxide to be removed. Future studies could explore the possibility of charge-controlled or strain-controlled CO$_2$ capture using beryllonitrene for easier regeneration. Additionally, the calculations performed in this study were only theoretical. In the future, these results can be applied to experimental carbon dioxide adsorbents. In order to maximize the abilities of beryllonitrene, future research could also be done to investigate the effects of vacancies and other dopants on the CO$_2$ capture ability of beryllonitrene.

\newpage

\bibliography{pu2021.bib}{}
\bibliographystyle{apsrev}

\end{document}